# Advances in multijunction solar cells: an overview

Anika Tabassum Raisa[a, b], Syed Nazmus Sakib[c], Mohammad Jobayer Hossain[d], Kaiser Ahmed Rocky[b, *], Abu Kowsar[a, *]

[a]*Bangladesh Council of Scientific and Industrial Research (BCSIR), Dhaka-1205, Bangladesh.*
[b]*Department of Physics, University of Dhaka, Dhaka-1000, Bangladesh*
[c]*Department of Information and Communication Engineering, Daffodil International University, Bangladesh*
[d]*Khulna University of Engineering & Technology (KUET), Khulna-9203, Bangladesh*

[*]Corresponding Author.
*E-mail Addresses*: apukowsar@gmail.com (A. Kowsar), kaiserrocky@du.ac.bd (K. A. Rocky)

**Abstract.** The advanced multijunction solar cell (MJSC) has emerged as a frontrunner in photovoltaic literature due to its superior photoconversion efficiency (PCE) owing to its complex fabrication procedure and high costs. This article aims to systematically review the advancements of III-V MJSCs by focusing on computational modelling and experimental fabrication methodologies. In addition, it addresses the technical barriers that have hindered the progression of MJSC technology while also evaluating the current status and prospects of these cells. The findings indicate that III-V MJSCs hold significant promise for space applications. However, advancements in materials science, growth techniques, and structural optimization are crucial for reducing fabrication costs to make these cells more viable for terrestrial use. In this context, alternatives such as perovskite/Si or perovskite/chalcogenide tandem solar cells emerge as viable options. By synthesizing insights from a thorough analysis of recent literature, this review serves as a valuable resource for researchers, industry practitioners, and newcomers seeking to deepen their understanding of the research methodologies, growth techniques, and the associated challenges and opportunities within the realm of MJSCs.

**Keywords-** Multijunction solar cell, photovoltaic, III-V material, tandem, fabrication procedures.

## 1. Introduction

Single-junction solar photovoltaic (PV) cells convert sunlight into electricity by absorbing wavelengths up to a specific limit determined by their bandgap [1]. As a result, only a fraction of the solar spectrum can be efficiently utilized for energy conversion [2]. To maximize the utilization of the solar spectrum, the concept of multijunction solar cells has emerged, as illustrated in Fig. 1 [3]. MJSCs are heterostructure optoelectronic devices composed of multiple semiconductor sub-layers stacked on a substrate. They have a photoconversion efficiency potential of up to 86.4% [4], as they can utilize a broader range of solar irradiance, ensuring a wide photo-response [5]. Historically, MJSCs have been exclusively used in space applications [6]. However, they have also been used in terrestrial applications in both unconcentrated and high solar-concentration forms in recent years [7]. III–V semiconductor materials on silicon cells are going to be an alternative and attractive pathway, reducing the manufacturing costs and eventually incorporating this high-efficiency cell technology into the widely used flat-plate silicon PV [8].

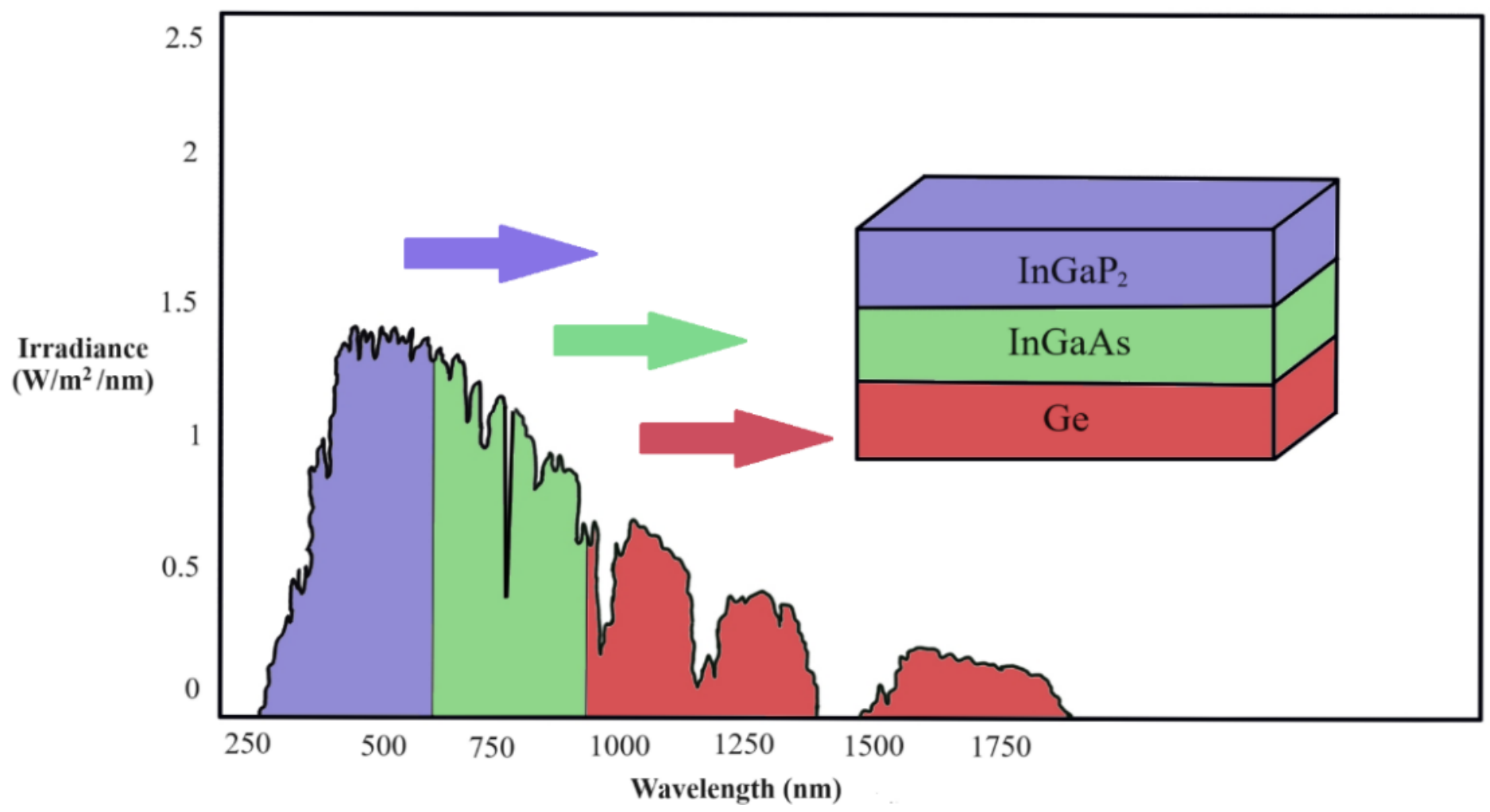

Fig. 1. Absorption of solar irradiation in different layers of a multijunction solar cell, with a triple-junction InGaP/InGaAs/Ge cell shown as an example [3].

Multiple p-n junction semiconductor sub-layer solar cells have been developed to capture a broader range of solar wavelengths, surpassing the Shockley-Queisser photoconversion efficiency limit of conventional PV cells [8]. The concept of MJSCs dates back to 1955 when semiconductor layers with varying bandgaps were stacked, with the highest bandgap material placed at the top to absorb shorter-wavelength photons [9, 10]. Sunlight passes through the topmost cell in the MJ structure, with each layer selectively absorbing photons within its energy bandgap range while allowing longer wavelengths to transmit through to lower-bandgap layers [11, 12]. This MJ approach helps reduce thermalization losses, which occur when high-energy photons are absorbed by low-bandgap materials, as well as below-bandgap losses, where low-energy photons fail to excite electrons in high-bandgap materials [9, 10]. These achievements are attributed to extensive research and development efforts since the late 1970s, along with advancements in bandgap engineering, high-quality epitaxial growth, and lattice matching [11].

**Table 1.**
Acronym and frequently used terms.

| Acronym | Nomenclature |
|---|---|
| AM | Air Mass |
| ARC | Antireflection Coating |
| CdTe | Cadmium Telluride |
| CIGS | Copper Indium Gallium Selenide |
| CZTS | Copper Zinc Tin Sulfide |
| COMSOL | Computer Solution |
| CPV | Concentrator Photovoltaic |
| DSSC | Dye-Sensitized Solar Cell |
| EQE | External Quantum Efficiency |
| FF | Fill-Factor |
| LPE | Liquid-Phase Epitaxy |
| $J_{SC}$ | Short Circuit Current Density |
| MBE | Molecular Beam Epitaxy |
| MOCVD | Metal-Organic Chemical Vapor Deposition |
| MOVPE | Metal-Organic Vapor Phase Epitaxy |
| MJSC | Multijunction Solar Cell |
| MPPT | Maximum Power Point |
| OMVPE | Organometallic Vapour Phase Epitaxy |
| PCE | Photoconversion Efficiency |
| PV | Photovoltaic |
| PVA | Polyvinyl Alcohol |
| PSC | Perovskite Solar Cell |
| $V_{OC}$ | Open Circuit Voltage |
| SiC | Silicon Carbide |
| TPSCs | Tin-based Perovskite Solar Cells |

The semiconductor materials in an MJSC structure are typically connected in series with ohmic contacts to ensure efficient operation, as individual junctions generate unequal voltages [12]. A significant advancement occurred in 1988 with a double heterostructure GaAs tunnelling junction, facilitating electron recombination with minimal energy loss and achieving 20% efficiency [13, 14]. The efficiency was improved by double hetero-wide bandgap tunnel junctions [15], hetero-face structure bottom cells, precise lattice-matching, and metamorphic cells [16]. Implementing antireflection coating and classifying the process to achieve the desired bandgap is also helpful in the efficiency increment of the next-generation MJSCs [17]. Polymer quadruple MJSCs with optimised materials and device structures may assist in achieving higher efficiency [18]. The most widely used multijunction combination is the triple-junction MJSCs, comprising three semiconductor sub-layers connected by tunnelling junctions [19]. Subsequent improvements led to 5-junction and 6-junction MJSCs, achieving impressive efficiencies of 35.8% and 47.1%, respectively [20, 21]. As a result, MJSCs have become integral to space exploration missions and hold promise for terrestrial concentrated photovoltaic systems, addressing energy demands and reducing carbon emissions. Looking ahead, Si-based double-junction tandem cell combinations, such as III–V/Si, II-VI/Si, chalcopyrite/Si, CZTS/Si, and perovskite/Si cells, are anticipated to play a crucial role in achieving highly efficient and cost-competitive photovoltaic cells for commercial manufacturing [22-26]. Additionally, other approaches like perovskite/perovskite, III–V/CIGSe, and perovskite/CIGSe MJ solar cells are still in the early stages but hold potential as candidates for future photovoltaic energy conversion [11, 27-29]. Nevertheless, a comprehensive literature review is undertaken to comprehend the current state-of-the-art MJSCs. Table 2 demonstrates the research focuses of the relevant studies, key discoveries, and knowledge gaps.

**Table 2.**
Review articles on multijunction solar cells till date.

| Authors | Review Focus | Summary of Findings | Knowledge Gaps |
|---|---|---|---|
| Yamaguchi et al. 2005 [16] | Review the present status and future potential of III-V MJSC up to 2005. | ▪ PCE of double junction and triple junction MJSCs were presented, mentioning record 37.4% PCE was achieved for InGaP/InGaAs/Ge cell with 200 suns.<br>▪ Future prospect of super-high PCE of concentrator MJSCs were also discussed. | Subcell materials, and cost reduction of the MJSC fabrication process were not addressed in details. |
| King et al. 2007 [30] | Progress of the high-efficiency III-V MJSCs. | ▪ Two high-efficiency MJSC structures were demonstrated, marking the first solar cells to surpass the 40% milestone.<br>▪ Promising future for concentrator PV technology. | The complexities and challenges were not addressed in the case of fabricating four junction MJSCs. |
| Baur et al. 2007 [31] | The viewpoints and obstacles related to the market integration of MJSC. | ▪ The implementation of III-V MJSC concentrator systems for advancing the economic feasibility of solar energy.<br>▪ The MAPCON system, collaborative efforts, and ongoing research for commercialization. | The authors focussed only on the III-V semiconductors. Other materials were not explored. |
| Friedman 2010 [17] | The advancements and challenges associated with material characteristics of MJSCs | ▪ Lattice-mismatched materials for achieving desired bandgaps.<br>▪ Anticipating the demonstration of 45% efficiencies in the near term and the realistic goal of approaching 55% efficiencies in the longer term. | The adverse impacts of lattice mismatch conditions were not adequately explored. |
| Siddiki et al. 2010 [18] | Polymer MJSCs with higher efficiency for large-scale application. | ▪ The progresses of polymer MJSCs had summarised up to 2010.<br>▪ Estimated the theoretical PCEs potential up to 24% for 4J cell. | It focused only on the polymer MJSCs and missed the exploration of organic-inorganic multijunction combinations. |
| Yamaguchi et al .2017 [32] | PCE obtaining potential for conventional and emerging MJSCs. | ▪ Performance comparison of various solar cells of NEDO project and the route to achieving solar cells with enhanced efficiency. | Defect behaviour, improved passivation on the front, rear, and interface surfaces, and optimization of series and shunt resistances were not fully addressed. |
| Colter et al. 2018 [19] | Incorporation and performance analysis of tunnel junctions for III-V MJSCs | ▪ AlGaAs/GaAs structure demonstrated the highest conductance with equivalent doping.<br>▪ A deterioration in tunnelling current has been noted irrespective of the material system. | It focuses only on the integration and impact of tunnel junctions in MJSC, especially AlGaAs/GaAs tandem cells. |
| Li et al. 2021 [33] | Strategies for enhanced stability of mixed-halide wide bandgap perovskite solar cells. | ▪ A high-quality film characterized by large grains, high crystallization and reduced defect density can be obtained by involving various processing conditions and strategies. | The mechanism of radiation damage in MJSCs with diverse materials and structures is not fully comprehended. |
| Wiesenfarth et al. 2018 [34] | Investigate the challenges in designing CPV | ▪ The impact of active-passive and thermal heat distributor.<br>▪ Encapsulation design in case of reduced module height and shorter focal distance.<br>▪ Reliability and usage of diffusion irradiation. | Overcoming chromatic aberration in primary and secondary optical elements requires further research and understanding. |
| Yamaguchi et al. 2021 [35] | Review the highly efficient MJSCs. | ▪ Reassessing MJSCs in terms of efficiency, cost-effectiveness, and potential market applications, considering scientific technological aspects. | Different recombination models in MJSC need to be explored, and practical strategies to reduce that recombination are yet to be identified. |
| Baiju et al. 2022 [36] | Review the multijunction combination from cell level to module for space and CPV applications. | ▪ MJSCs with enhanced performance through research and development efforts.<br>▪ Innovative concepts and materials aiming to enhance efficiency by minimising thermalization and transmission losses. | The manufacturing cost reduction process of this technology was not clearly addressed. |
| Verduci et al. 2022 [6] | The utilization of solar photovoltaic energy and technologies for space applications | ▪ Reviewed and summarized commercially available (Si-and MJSCs) and emerging (CIGS and perovskite) solar cells for space applications. | Implementing radiation resistance in solar arrays while maintaining cost-effectiveness is an area that needs further exploration and study. |

Table 1 presents a compilation of abbreviations and terminology used in this work. Table 2 presents a summary of relevant articles from various journals covering different aspects of MJSCs, including their status, limitations, efficiency enhancements, terrestrial and space applications, and prospects. A thorough examination of these reviews and related literature has highlighted several research gaps. While MJSCs have shown significant potential for space and terrestrial applications, further advancements are needed, particularly in materials science, growth techniques, and structural optimization. Additionally, the complexity and production costs of MJSCs remain higher than those of commercially produced silicon solar cells. Addressing these challenges requires a deeper exploration of critical factors such as material growth, fabrication strategies, multi-junction configurations, and emerging trends. This review aims to comprehensively analyse these key aspects, which have received limited attention in previous studies. The specific objectives of this review are to systematically examine the progress of MJSCs in the following areas:

- Advancements in MJSCs with two to six subcells.
- Computational approaches driving MJSC research forward.
- Fabrication procedures enhance efficiency and reduce production costs.
- Materials for MJSCs, including III-V compounds, hybrid tandems, and emerging materials.
- Impact of irradiance concentrations on MJSC performance.
- Applications and cost feasibility of MJSCs.

This review explores future research directions in III-V MJSC, highlighting critical areas that require further investigation to bridge the gap between laboratory demonstrations and practical deployment. To the best of the authors' knowledge, a comprehensive review article rigorously covering these topics is rare. This paper is structured as follows: Section 2 details the materials and methodologies employed in this study; Section 3 provides a critical analysis of the latest advancements in MJSC technology, aligned with the specified objectives; Section 4 evaluates the economic viability of MJSCs; Section 5 offers an in-depth discussion and future perspectives; and Section 6 concludes with key insights and final remarks.

## 2. Materials and methods

This article conducts a comprehensive review of over 160 peer-reviewed publications, focusing on the significant advancements in MJSC technology. This review encompasses a variety of scholarly outputs, including original research articles, critical review articles, and conference proceedings, all authored by distinguished experts in the field of advanced photovoltaics [37]. The publications analyzed in this study were meticulously sourced from reputable research databases, such as Web of Science (SCI, SCIE and ESCI), Scopus, ScienceDirect, the DOAJ, IEEE Xplore, and Google Scholar. Notably, a substantial portion—near 150—of these reviewed works were retrieved from the Scopus database. This particular collection is noteworthy not only for its volume but also for its high h-index scores, which serve as an indicator of the quality and impact of the articles. A high h-index suggests that other researchers have frequently cited these publications, reinforcing their significance and value within the renewable energy sector. Through this extensive literature review, the article aims to illuminate the current state of research and identify pivotal trends and breakthroughs in the domain of multijunction solar cells.

## 3. Review of multijunction solar cell

The following subsections discuss the advancements in subcell combinations of MJSCs, their lattice matching and mismatching conditions, computational approaches, fabrication procedures, and a comparison with polymer tandem applications.

### *3.1. Subcell layers of MJSC*

MJSCs began with two-junction (2J) AlGaAs/GaAs solar cells [41], where a tunnelling diode connected two subcells in series. The open-circuit voltage ($V_{oc}$) of the 2J cell was estimated to be 2.0 V, and the authors also demonstrated the I-V characteristic curve. Fabricating a monolithic cascaded structure requires further optimization to enhance the PCE of the solar cell. In addition to the III-V combination, an article by Bailie and McGehee [42], published in 2015, demonstrated the use of metal-halide perovskites and solution-processable large-bandgap materials in the 2J solar cell model. However, the instability of mixed-halide compounds presents a challenge to achieving optimal 2J cell efficiency. The highest reported PCE for a state-of-the-art silicon-perovskite tandem is 33.7% [38]. However, the theoretical PCE of double-junction cells was estimated at 46.1%, significantly higher than that of single-junction cells with a PCE limit of 33.7%. To date, the highest PCE in the case of dual-junction III-V solar cell was reported as 35.5% under a concentration of 38 suns, which was achieved by fabricating a lattice-mismatched GaInAsP/GaInAs solar cell grown by atmospheric-pressure organometallic vapour phase epitaxy (OMVPE) [39].

Geisz et al. [40] monolithically developed a three-junction (3J) InGaP/GaAs/InGaAs solar cell using OMVPE with a $V_{oc}$ over 2.95 V and PCE of 33.8%. The Ge-free inverted configuration was chosen to make it cost-effective and more efficient. The advancement of triple-junction AlGaAs/GaAs/InGaAs solar cells with an anti-reflective coating based on a double layer $MgF_2/ZnS$ was reported in [41] with $V_{oc}$ of 3.2 V and a PCE of 38.5%. Sasaki et al. [42] developed a 43.5% efficient InGaP/GaAs/InGaAs triple-junction solar cells using an inverted configuration grown by metal-organic chemical vapor deposition (MOCVD). To date, the highest reported PCE (44.4%) for 3J solar cell was obtained under 302 suns [43, 44]. A 4J solar cell was introduced by Riesen et al. [45], where the subcells of the MJSC were current-matched, and the module design was optimised to reduce cost and optical transmission losses with an estimated PCE of 38.9%. The fabrication and analysis of wafer-bonded 4J GaInP/GaAs/GaInAs/GaInAsSb MJSC grown using OMVPE were reported in [46], with 42% PCE under 599 suns condition under AM1.5D condition. However, the state-of-the-art highest efficiency 4J GaInP/GaAs/GaInAsP/GaInAs solar cell was fabricated and reported with 46% PCE [47]. Zhang et al. reported a five junction (5J) AlGaInP/AlGaInAs/GaAs/GaInNAs/Ge solar cells using AlGaInP/Ge 2J solar cells grown by MOCVD with Sb incorporation [48]. However, this cell structure provided a 15.3% higher Jsc, overcoming the problems of direct bonding. A theoretical demonstration has been performed for an InP-based 5J InGaP/InGaP/InGaAsP/InGaAsP/InGaAs MJSC was investigated in [49]. The current-matched 5J model shows an ideal PCE of 53.9% under 1000 suns condition. Alternatively, a direct semiconductor-bonded 5J solar cell has been reported to achieve the highest PCE of 38.8% (non-concentrating condition), which was grown using MOVPE [20]. MJSC research further advances to a six-junction (6J) combination to enhance the PCE; therefore, Geizs et al. fabricated an inverted metamorphic 6J structure with a PCE of 35.8% [50]. Geizs-led same group further reported the ever-highest record 47.1% PCE in the photovoltaic literature for another 6J cell combination for inverted metamorphic concentrating condition with $V_{oc}$ of 5.15V [51].

**Table 3.**
A comparative study of notable MJSCs (Y: year of publication, N: nature of the work, J: number of junctions, $J_{SC}$: short circuit current density, $V_{OC}$: open circuit voltage, FF: fill factor, $\eta$: efficiency).

| Author | J | Y | Solar Cell | N | Tool | $J_{SC}$ ($mA/cm^2$) | $V_{OC}$ (V) | FF (%) | $\eta$ (%) |
|---|---|---|---|---|---|---|---|---|---|
| Bedair [52] | 2 | 1979 | AlGaAs/GaAs | Experimental | LPE | 7 | 2.0 | 70-80 | 25 |
| Bertness [53] | 2 | 1994 | GaInP/GaAs | Experimental | APOVPE | 14 | 2.99 | 88.5 | 29.5 |
| Jain [39] | 2 | 2018 | GaInAsP/GaInAs | Experimental | MOVPE | 18.59 | 2.23 | 85.7 | 35.5 |
| Tiwari [54] | 3 | 2016 | GaP/InGaAs/InGaSb | Theoretical | - | 7.4 | 3.32 | - | 23.5 |
| Geisz [40] | 3 | 2007 | InGaP/GaAs/InGaAs | Experimental | OMVPE | 13.1 | 2.95 | 86.9 | 33.8 |
| Correa [41] | 3 | 2015 | AlGaAs/GaAs/InGaAs | Theoretical | - | 13.7 | 3.2 | 90.0 | 38.5 |
| Sasaki [42] | 3 | 2013 | InGaP/GaAs/InGaAs | Experimental | MOCVD | - | 3.01 | 86 | 37.7 |
| Predan [46] | 4 | 2019 | GaInP/GaAs/GaInAs// GaInAsSb | Experimental | OMVPE | 12.19 | 4.09 | 84.2 | 42.0 |
| Dimroth [47] | 4 | 2015 | GaInP/GaAs//GaInAsP/GaInAs | Experimental | MOVPE | - | 4.23 | 85.1 | 47.6 |
| Zhang [48] | 5 | 2017 | AlGaInP/AlGaInAs/GaAs/GaInNAs/Ge | Experimental | MOCVD | - | - | - | - |
| Huang [49] | 5 | 2015 | InGaP/InGaP/InGaAsP/InGaAsP/InGaAs | Theoretical | - | 10.65 | 4.24 | 87.0 | 43.6 |
| Geisz [50] | 6 | 2018 | GaInAs/ GaInAs/ GaInAs/GaAs/AlGaAs/AlGaInP | Experimental | MOVPE | 8.05 | 5.30 | 83.9 | 35.8 |
| Geisz [55] | 6 | 2020 | AlGaInP/AlGaAs/GaAs/GaInAs(3) | Experimental | OMVPE | 5.6 | 5.15 | 86 | 47.1 |

The incremental trends of the photoconversion efficiency of MJSC are presented in Fig. 2.

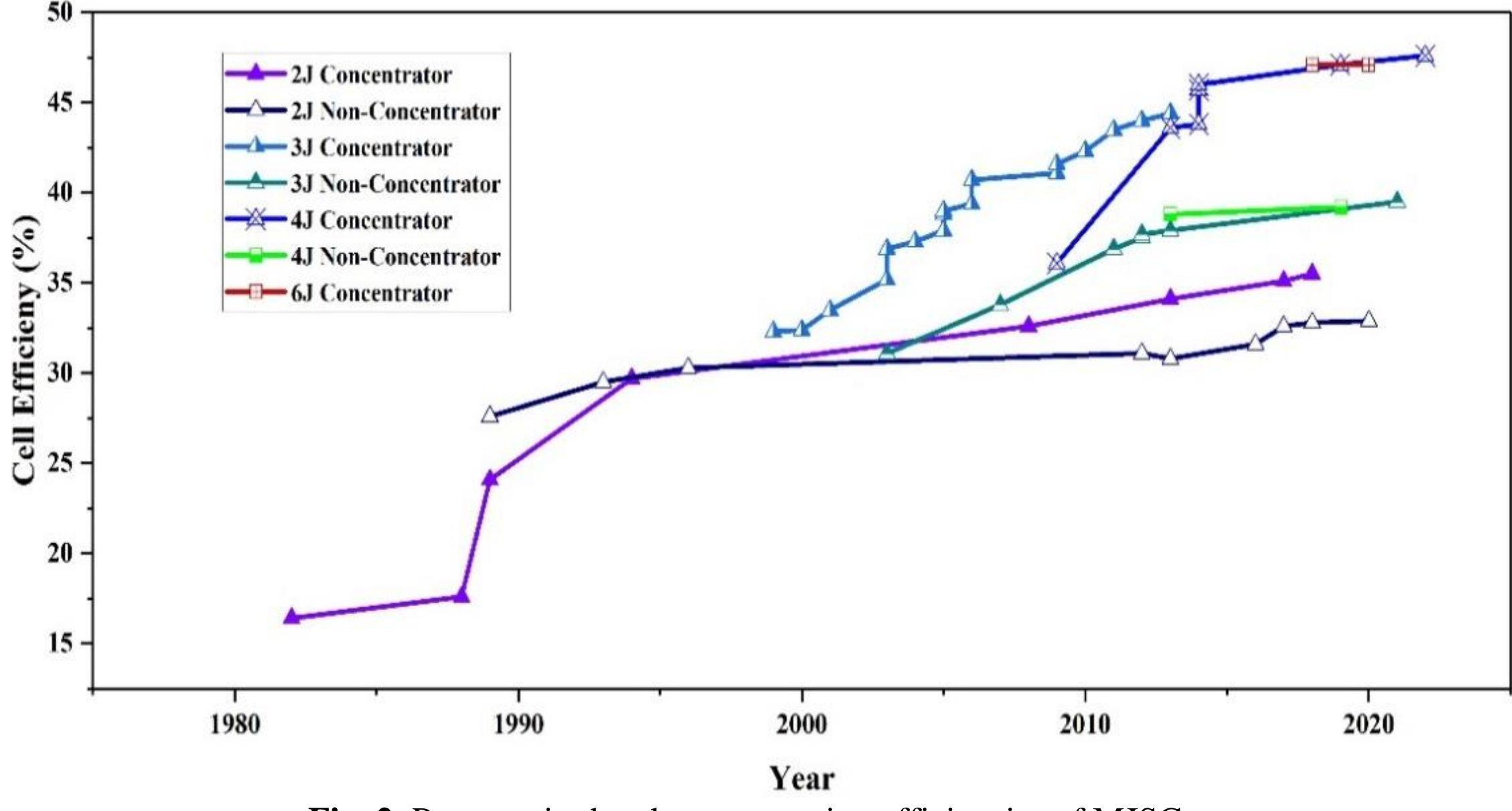

**Fig. 2.** Progress in the photoconversion efficiencies of MJSCs.

Nevertheless, reducing the series resistance within the subcell of the 6J cell structure could further enable the realization of the PCE over 50% [55] and, as a consequence, advance MJSC research to enhance efficiency and reduce fabrication cost using both theoretical and experimental approaches, using different cell combinations with different materials for spacecraft as well as terrestrial applications. The champion efficiencies for 2-6 junction configurations have been summarized in Table 3.

### *3.2. Active Materials in MJSCs*

MJSCs consist of two or more stacked subcells, each made of semiconductor materials with carefully chosen bandgaps in order to optimize energy conversion. The development of MJSC growth materials has been driven by the need to maximize photon absorption, minimize recombination losses, improve stability under various environmental conditions, and enhance overall PCE [11, 36, 56]. Over the years, researchers have systematically introduced new materials to overcome the limitations of earlier technologies, leading to significant improvements in performance. The III-V semiconductor group has played a fundamental role in MJSC development due to its tunable bandgap, high carrier mobility, and superior optoelectronic properties. Materials such as indium gallium phosphide (InGaP), gallium arsenide (GaAs), indium phosphide (InP), aluminium gallium arsenide (AlGaAs), indium aluminium arsenide (InAlAs), gallium indium arsenide phosphide (GaInAsP), gallium indium nitride phosphide (GaInNP), indium gallium arsenide (InGaAs), gallium arsenide bismide (GaAsBi), and germanium (Ge) have been widely used as absorbers in different subcells of MJSCs [8, 57-61]. GaAs has been particularly important due to its direct bandgap of 1.42 eV, which allows for efficient photon absorption and carrier transport [9]. However, GaAs-based solar cells suffer from high surface recombination rates, leading to efficiency losses. To mitigate this issue, researchers introduced window layers on top of the GaAs surface, which helped passivate surface defects and reduce recombination losses [62, 63]. Despite these improvements, external factors such as temperature, atmospheric pressure, impurities, airmass variations, and radiation exposure continue to affect the performance of GaAs-based MJSCs.

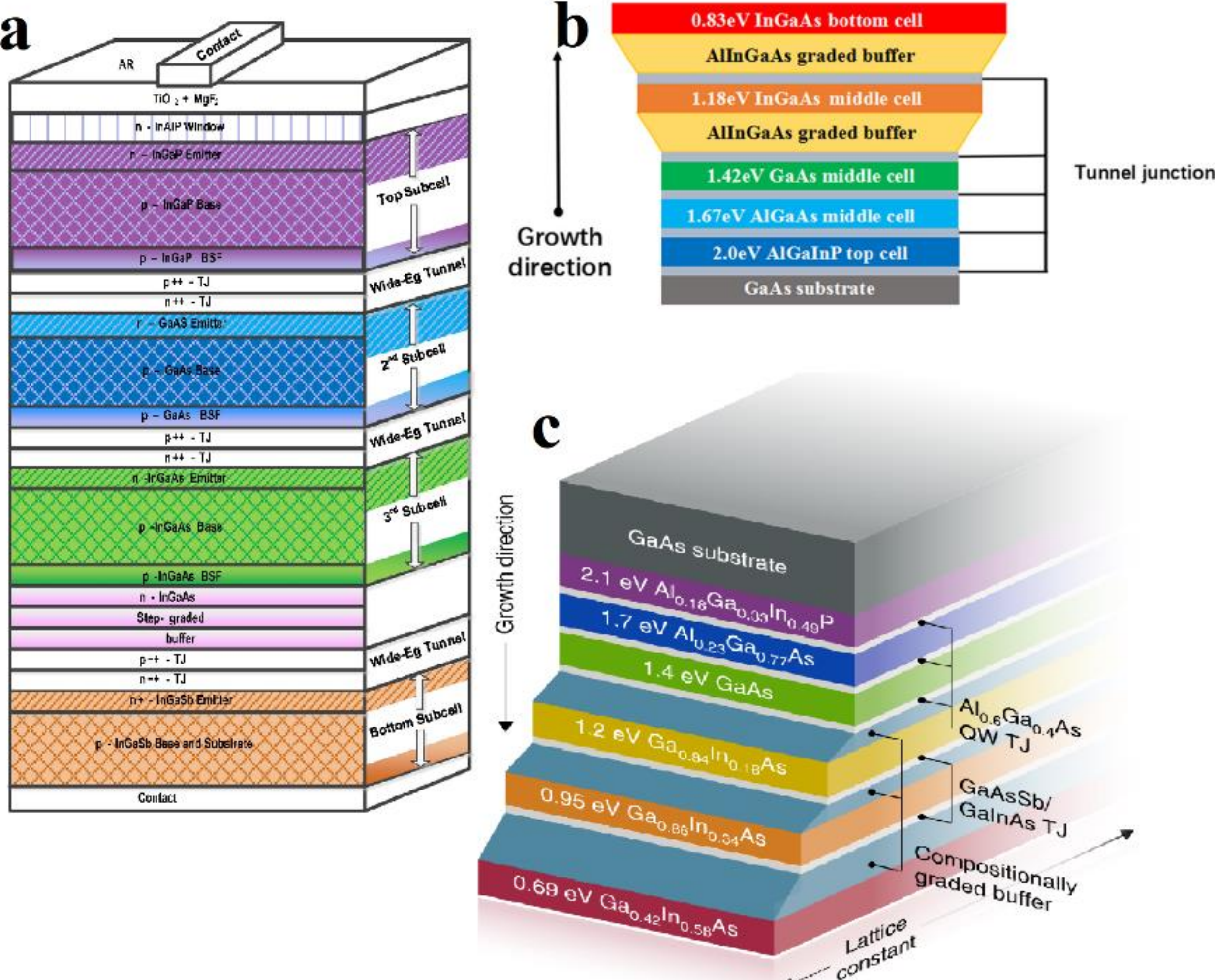

Fig. 3. (a) 4J cell [64], (b) flexible and large-size inverted metamorphic 5J cell [65], (c) champion 6J cell [55].

Papez et al. [66] investigated the effects of gamma radiation on GaAs-based solar cells using synthetic radioactive Co-60 and found that the solar cells remained operational even after exposure to doses as high as 500 kGy. However, they also observed that at extreme radiation doses, the electrical parameters of the solar cell began to deteriorate. Similarly, Feteha [67] reported a reduction in PCE and FF under prolonged radiation exposure, emphasizing the need for materials with higher radiation resistance for applications in space and extreme environments.

Researchers explored stacking GaAs with other III-V materials to improve MJSC efficiency and overcome the limitations of single-junction solar cells. Tandem and multijunction configurations were developed to maximize absorption across a wider range of the solar spectrum [11, 68]. A mechanically stacked III-V and silicon (Si) tandem cell achieved an efficiency of 32.8% [23]. However, the increment of efficiency of 35.9% was achieved by incorporating a GaInP/GaAs stack with a Si single junction cell [23]. As a consequence of the advancement, a 6J inverted metamorphic structure, mentioned earlier, has shown an efficiency of 39.2% under normal global (1 sun) illumination [55]. Concentrator photovoltaic (CPV) technology further expanded the efficiency potential of MJSCs. By integrating a reverse heterojunction AlGaInP subcell, researchers were able to achieve a peak efficiency of 47.1% under 143 suns (AM1.5D conditions), demonstrating the impact of high-concentration sunlight on multijunction configurations. It also suggested that reducing the series resistance in this structure even further could potentially push efficiencies beyond 50% [69]. Fig. 3 illustrates solar cell models with 4, 5, and 6 junctions.

Beyond III-V materials, hybrid solar cell approaches, particularly those involving perovskite-silicon tandems, have gained increasing attention due to their cost-effectiveness and high absorption capabilities. Recent studies have demonstrated the potential of organic-inorganic perovskites to complement traditional semiconductor materials [70, 71]. Sun et al. incorporated a water-soluble addition called polyvinyl alcohol (PVA), which is both cost-effective and widely available. The inclusion of PVA was reported to result in a 17.4% *PCE*, an 11.6% increase compared to devices without the additive. Furthermore, the PSCs with PVA were reported to maintain more than 90% of their initial efficiency even after operating in a high-humidity environment for 30 days. To increase the effective *PCE*, perovskite absorbers with an ideal bandgap (1.3-1.4 eV) were introduced [72]. An improved *PCE* (up to 17.63%) was achieved by designing a new absorber composition.

Despite the promising advancements in perovskite solar cells, the presence of lead (Pb) in traditional PSCs poses environmental concerns due to potential Pb leakage. Addressing this challenge, researchers have explored various encapsulation techniques to prevent lead outflow from damaged devices [73]. Jiang *et al.* [74] used $Pb^{2+}$-absorbing materials for physical encapsulation, effectively preventing the outflow of lead from damaged devices. A chemical approach was incorporated using on-device sequestration of Pb leakage by applying Pb-absorbing thin-films on both sides of the multijunction cell stack [75]. Another approach to solving the problem of Pb leakage is to use tin-based PSCs (TPSCs) as Pb-free alternatives to traditional PSCs. A fabrication method for TPSC was introduced that achieved a stabilised efficiency of 11.22% and, after operating at the maximum power point (MPPT) for 1000 hours, maintained more than 95% of their initial efficiency [76]. These developments highlight the ongoing efforts to enhance both the performance and environmental sustainability of perovskite-based multijunction solar cells. A major breakthrough was achieved in monolithic perovskite/silicon tandem cells, where a two-junction (2J) perovskite/Si device demonstrated a record-breaking efficiency of 31.3% for a one-square-centimeter cell, surpassing the long-standing 30% milestone [77].

In addition to III-V and perovskite-based materials, nitride-based solar cells have shown great potential for multijunction applications [78-81]. Nitride compounds such as gallium indium nitride arsenide antimonide (GaInNAsSb) have been explored for their ability to enhance photon absorption and carrier transport in MJSCs. A GaInP/GaAs/GaInNAsSb/GaInNAsSb four-junction solar cell, grown using MBE, achieved an efficiency of approximately 39% [82]. Other nitride-based materials, including dilute GaAsSbN layers and p-i-n heterostructures grown by liquid phase epitaxy, demonstrated efficiencies of around 4.1%, showing the potential of these materials in specialized applications [83]. Besides the standard materials, a few emerging materials and combinations show their potential to become a good alternative [84, 85]. Various tandem structures such as copper tin zinc sulfide (CTZS)/Si [86], n-type cadmium sulfide (n-CdS)/p-type copper indium gallium selenide (p-CIGS)/p+-copper gallium selenide (p+-CGS) [87], and Au/Spiro-OmeTAD/CIGS/MASnI3/CdS/ZnO/FTO [88] solar cells have been proposed as potential candidates for next-generation photovoltaic technologies. While these materials have demonstrated encouraging results in early-stage experiments, further validation and comprehensive techno-economic assessments are necessary to determine their viability for large-scale manufacturing and commercial deployment.

The continuous advancement of MJSC materials has led to significant improvements in efficiency, stability, and environmental sustainability. The transition from traditional III-V materials to hybrid perovskite-based tandems and nitride-based architectures demonstrates the dynamic evolution of solar cell technology. While perovskite/silicon tandems have set new efficiency benchmarks, challenges such as lead leakage and long-term stability must be addressed through innovative encapsulation techniques and alternative material choices. Emerging materials such as CTZS and CIGS-based tandem structures offer promising directions, though further experimental validation is required. Moving forward, optimizing material selection, refining stacking architectures, and developing robust encapsulation strategies will be key to achieving the next generation of highly efficient and commercially viable MJSCs.

### *3.3. Lattice matching and mismatching conditions of MJSC*

MJSCs have been extensively developed based on lattice matching and mismatching growth techniques, both of which play a critical role in determining the efficiency and structural integrity of the solar cell. Lattice matching refers to the alignment of the crystal structures of semiconductor materials within the different cell layers, ensuring minimal strain at the interfaces. This structural coherence allows for optimal electron movement, reducing recombination losses and preventing defect formation, which could otherwise degrade performance [36, 89]. In contrast, lattice mismatching occurs when the crystal structures of different layers do not align perfectly, leading to the formation of dislocations and defects. While improper management of these defects can severely impact photovoltaic performance, lattice mismatching can also be strategically utilized to alter the bandgap of the subcells, enhancing photon absorption and thereby increasing PCE [90]. The choice between lattice matching and mismatching ultimately depends on the specific requirements of the MJSC, with both approaches offering distinct advantages and challenges.

Early advancements in lattice-matched MJSCs were demonstrated in III-V tandem solar cells, such as those incorporating a GaNPAs top cell, a sandwich GaP-based tunnel junction (TJ), and a diffused Si-bottom cell grown epitaxially on a silicon substrate [91]. In the absence of an antireflection coating (ARC) layer, the cell attained a $V_{oc}$ of 1.53 V and a PCE of 5.2% under the AM1.5G global illumination condition. To achieve higher efficiencies, further advancements in the upper junction and connecting junction performance are necessary, requiring improvements in various aspects of material growth and doping control. The efficiency limitation in the lattice-matched 3J solar cells was successfully overcome by adopting the system of dilute nitride materials [92]. These lattice-matched concentrator cells have achieved independently verified efficiencies of 43.5% by NREL and Fraunhofer. Continuing the trend, a monolithic 3J lattice-matched InAlAs/InGaAsP/InGaAs solar cell with an optimised band gap combination was modelled for three subcells having specific band gaps 1.93, 1.39, and 0.94 eV, respectively, and lattice constant of 5.807 A° [93]. The device simulations showed that the proposed approach was able to achieve theoretical PCEs above 51% under 100-sun illumination, indicating its potential for high efficiency. These developments indicate the potential of lattice-matched MJSCs for further efficiency improvements through refined material engineering. Recent advancements in thin-film 3J cells have been realized using metal-organic chemical vapour deposition (MOCVD) techniques in InGaP/(In)GaAs/Ge structures [94] and for 4J GaInP/GaAs/GaInNAsSb/GaInNAsSb solar cell using molecular beam epitaxy (MBE) technique [82]. However, lattice-matched structures have been successfully employed in 2J–4J MJSCs, and the inverted metamorphic approach has emerged as a promising alternative for high-efficiency, next-generation MJSCs, offering greater flexibility in material selection.

The lattice-mismatching growth technique presents challenges as well as offers opportunities for MJSCs; however, it enables the use of a greater variety of materials, expanding the options for bandgap engineering. By intentionally introducing lattice-mismatched materials, tuning the bandgaps of individual junctions to match specific regions of the solar spectrum becomes possible. Geisz *et al.* proposed lattice-mismatched p-on-n GaAsP photovoltaic cells on Si achieved through a compositional step-graded buffer [95]. These cells perform on par with, or even better than, previous reports of solar cells using Triethylchloroarsine to grow AlGaAs and GaAsP on Si. Tandem solar cells comprised of InP/InGaAsP were studied to observe lattice-mismatch impact on efficiency [96]. Apart from material growth considerations, temperature variations and spectral shifts are a significant challenge in MJSC performance, particularly in outdoor applications. Lattice-mismatched MJSCs are more susceptible to thermal expansion mismatches, which can introduce additional stress at interfaces, leading to long-term degradation. In concentrated photovoltaic (CPV) systems, where MJSCs operate under extremely high solar intensities, heat dissipation becomes a critical issue, further complicating performance optimization. The impact of lattice mismatching on temperature resilience was investigated through the long-term outdoor performance evaluation of CPV systems, demonstrating the importance of thermal management in real-world conditions [97]. While lattice-mismatch techniques enable the use of a broader range of materials, they also introduce greater fabrication complexity and cost, which can be barriers to large-scale deployment. Besides the effective growth of MJSCs using the mismatch technique, the failure of this technique was also analysed by Long *et al.* [98]; however, fabricating MJSC using this approach creates more complexity, which, in turn, increases the cost of the solar cell. Moreover, to obtain ultra-high-efficiencies, the latest champion 3 to 6J solar cells utilize the lattice-mismatch technique with step-graded layers [55, 99-101]. As research continues to refine growth techniques and enhance thermal stability, lattice-mismatched MJSCs are expected to play a vital role in the next generation of high-performance photovoltaic technologies.

### *3.4. Computational approaches for designing MJSCs*

Since the advent of MJSCs, computational modelling has become influential in predicting device performance, optimizing material configurations, and advancing the development of high-efficiency solar cells. Achieving high PCE in MJSCs needs meticulous control over bandgap alignment, charge carrier dynamics, spectral response, and current matching between subcells. Due to the inherent complexities of several experimental processes associated with MJSCs, extensive numerical simulations have been developed to support device optimization prior to fabrication. These computational methods not only enable researchers to explore innovative material combinations and evaluate the impact of optical and electrical losses but also significantly decrease dependence on costly and time-consuming experimental trials, thereby restructuring the process and enhancing reliability [15, 102, 103]. Kurtz et al. demonstrated the theoretical models, focusing on two major categories: (a) detail balance model and (b) 1D transport model [104]. The detailed balance model provides an upper-efficiency limit by assuming ideal conditions, such as perfect absorption, negligible non-radiative recombination, and ideal current matching between subcells. In contrast, the 1D transport model incorporates charge carrier transport dynamics, considering factors such as recombination, diffusion, and electrical losses, thereby offering a more realistic performance prediction. She also elucidated the spectral responses with different approaches and PCEs for MJSCs. Nell and Bernett developed another classical model, the spectral p-n junction model, for 2J solar cells to numerically realise the efficiency under normal and concentrated solar conditions [105]. They have developed an expression for the efficiency improvement of increasing sun concentration. Based on the model, using MATLAB, Mouri et al. [3] have shown AlAs/GaAs/Ge-based solar cells can achieve an efficiency of 44.52%. Similar approaches have been demonstrated for various MJSCs with different materials, as sub-cells have shown great potential for higher efficiency without applying concentrator systems [2]. MJSC with the configuration AlAs/GaAs/$GaAs_{0.9}Bi_{0.085}$ can achieve 48% efficiency theoretically [106], $GaInP_2$/GaAs/$GaAs_{0.94}Bi_{0.0583}$/$GaAs_{0.91}Bi_{0.0857}$ shows PCE of 52.2% [58], $In_{0.1}Ga_{0.9}N$/SiC/Si can yield 60.07% PCE in numerical simulation [107]. Sakib et al. compared various solar cell parameters for different III-V MJSC through simulation, showing efficiency improvement with increasing subcells and solar concentrations [108]. These studies underscore the potential of computational approaches in predicting high-efficiency solar cell designs and guiding experimental innovations.

Despite their utility, computational models inherently rely on a set of assumptions that introduce discrepancies between simulated and experimentally realized efficiencies. Most numerical simulations assume idealized conditions such as negligible series and shunt resistances, minimal non-radiative recombination, perfect current matching, and uniform material quality [109-111]. These idealizations often result in higher predicted efficiencies compared to experimental demonstrations, where practical challenges such as fabrication defects, material inhomogeneities, optical losses, and thermal effects limit real-world performance. Moreover, computational models often exclude secondary factors such as degradation mechanisms, long-term stability concerns, and parasitic resistances that are crucial for practical implementation. As a result, while numerical studies provide a fundamental understanding of MJSC operation, experimental validation remains essential for translating theoretical predictions into commercially viable technologies.

The computational approaches are performed based on in-house coding, in most cases using MATLAB [44, 112, 113]. In addition, a few modelling tools, viz, SCAPS-1D, wxAMPS, and TCAD, could numerically simulate the performance metrics of MJSCs [114]. The Silvaco ATLAS tool has been utilized for numerical simulation and optimization of InGaP/GaAs/InGaAs/Ge four junction cells using the current matching conditions [115]. Besides the established simulators, MSCS-1D, a recent tool specially developed for MJSC, could effectively simulate the performance parameters [2, 116]. However, given the limitations of conventional computational models, integrating machine learning algorithms and genetic optimization techniques into MJSC simulations has shown promise in refining efficiency predictions, optimizing material selection, and accelerating the discovery of novel semiconductor configurations [117-119].

### *3.5. MJSC Fabrication Processes*

Multiple semiconductor layers with various bandgaps are piled onto a substrate to create MJSCs, which absorb near-ultraviolet to mid-infrared wavelengths of the solar spectrum, thereby increasing the solar cell's overall efficiency. Several techniques were found effective in fabricating MJSCs, and the following subsections critically analysed some commonly used techniques.

The Epitaxial Growth technique involves growing semiconductor layers on top of each other using epitaxial growth methods such as MOCVD [120], MOVPE [121], MBE [122], or liquid-phase epitaxy (LPE) [123]; among them, MBE is the commonly utilized method for fabricating III-V MJSCs [124]. For example, a high-quality 3J InGaP/(In)AlGaAs/GaAs solar cell was fabricated using solid-source MBE [125]. Sun *et al.* also fabricated solar cells incorporating AlGaInP by MBE and found an enhanced *PCE* compared to MOVPE [126]. The rapid thermal annealing process enhanced overall performance, allowing the 2.0 eV MBE-grown cells to achieve efficiency comparable to MOVPE-grown cells. On the other hand, GaInP/GaInAs/Ge cell has been on a large scale in MOCVD reactors [127]; besides, TCO/Cd(Zn)S emitter interface of cadmium telluride (CdTe) solar cells have been fabricated using this MOCVD method [128]. The MOVPE technique is frequently used in fabricating III-V material-based MJSCs [55]. For example, the fabrication of InGaP with a rate of 30 µm/h paved the way for producing III-V MJSCs in the high-speed MOVPE reactor [129].

Tanabe reviewed several wafer bonding techniques for single- and multijunction solar cells [130]. Wafer bonding is another effective technique for joining subcells to fabricate the IMM MJSCs [131]. This is typically achieved using techniques like metal diffusion bonding or adhesive bonding; the champion 4J solar cell with PCE of 42% was developed using this wafer-bonding method [132], III-V growth on Si [131], III-V MJSC including (Al)GaAs, GaAs, InGaAs and InP semiconductors was growth using room temperature wafer bonding [133] are few classical reports. Besides the *PCEs*, Tayagaki et al. investigated the properties of different wafer-bonded semiconductor parameters [134].

Monolithic integration involves growing semiconductor layers on a single substrate [135]. Solar cells with several junctions may be grown monolithically integrated with lattice-matched materials without causing misfit dislocations [135]. Conceptually, monolithically grown II–VI MJSCs show a potential to achieve *PCE*s up to 44% under normal condition and 50% under 500 suns concentrating condition [136], and 6J solar cell simulated a *PCE* of 43% in space, and 52% under 240 suns illumination [137] and experimentally III-V subcells grown on Si produce 19.7% *PCE* [138], 30.2% [135], and etc. It's worth noting that the specific techniques used for MJSC fabrication can vary depending on the material systems and bandgap combinations chosen. Researchers and manufacturers continue to explore and develop new techniques for improved efficiency and cost-effectiveness of MJSCs.

### *3.6. Concentrating conditions of MJSC*

Concentrator photovoltaic (CPV) systems have undergone significant advancements in recent years, making them an increasingly viable option for photovoltaic power generation. CPV systems utilise optical devices, such as lenses or mirrors, to direct sunlight toward small, high-efficiency MJSCs. This concentrated sunlight increases PCEs compared to traditional flat-plate photovoltaic systems. Recent advancements in refractive optical elements have made using a prismatic cell cover with a domed Fresnel lens concentrator possible to prevent metallisation losses [139]. This has led to significant reductions in the required area and mass compared to conventional space photovoltaic systems [140]. MJSCs integrating with CPV technology have achieved impressive PCEs, reaching well over 40% in laboratory settings. The highest-ever PCE for a concentrator MJSC was reported to be 47.1% for a 6-J CPV cell arrangement [55]. However, advancements in tracking and cooling systems have improved the overall performance and reliability of CPV systems. High-precision solar tracking systems ensure that the concentrator optics are aligned with the sun's position throughout the day, maximising the amount of sunlight captured. Single- and dual-axis solar tracking systems are the two basic types that differ in the degree of freedom of movement [141]. Effective cooling mechanisms assist in maintaining optimal operating temperatures for the photovoltaic cells, enhancing their efficiency and longevity. Hence, several cooling technologies have been investigated, such as liquid immersion, water cooling, and microchannel heat sinks [142, 143]. For a concentration ratio exceeding 20 suns, passive cooling for linear concentrators was reportedly insufficient [143]. Recent developments and challenges in cooling methods for concentrated photovoltaic thermal systems were discussed in [144]. This article also explored the principles of advanced cooling systems for photovoltaic and concentrated solar modules, focusing on thermal considerations while utilizing nanotechnology and enhancing performance.

CPV systems are advancing in incorporating energy storage technologies, significantly improving photovoltaic power generation's reliability and efficiency. By integrating energy storage solutions, CPV systems utilizing MJSCs can effectively address the intermittency associated with solar energy, leading to more stable power outputs and minimizing energy losses attributed to light discarding [145]. This is particularly advantageous for CPV-integrated MJSC applications, as the inherent high conversion efficiencies of MJSCs under concentrated sunlight can be optimized further when combined with suitable energy storage systems. Furthermore, innovations in thermal management remain essential for sustaining the high-performance levels of MJSCs in CPV configurations. Since MJSCs are exposed to high-intensity solar radiation, the implementation of effective cooling strategies—such as liquid immersion cooling and microchannel heat sinks—is critical for maintaining their long-term efficiency and operational stability [142, 143]. Elevated temperatures can compromise junction performance and carrier transport dynamics without adequate cooling, ultimately reducing PCE. Consequently, integrating precision tracking systems, advanced cooling mechanisms, and energy storage solutions within CPV-MJSC systems is vital for maximizing efficiency, prolonging operational lifespan, and facilitating widespread adoption in large-scale solar power ventures. These technological innovations bolster the practical applicability of CPV-integrated MJSC technology, positioning it as a competitive alternative to traditional photovoltaic systems.

### 3.7. MJSC application areas

The III-V MJSCs have been extensively explored for both terrestrial and space power generation applications due to their high-power conversion efficiencies. However, their adoption has been significantly more prominent in space applications, while their use in terrestrial environments remains limited. The primary reason for this disparity is the high manufacturing cost and complex fabrication process associated with III-V MJSCs, making them less economically viable for large-scale terrestrial deployment. Despite their superior efficiency compared to conventional silicon-based solar cells, the per watt cost remains significantly higher, restricting their widespread adoption for general commercial or residential use.

In terrestrial applications, MJSCs have primarily been utilized in CPV systems, where they operate under high solar concentration, often in the range of 500–1000 suns [146, 147]. CPV technology aims to enhance efficiency while minimizing the amount of semiconductor material required per unit area. However, this approach introduces additional challenges, such as the necessity for precise solar tracking systems to ensure optimal alignment with direct sunlight. The implementation of tracking mechanisms increases both the initial installation costs and ongoing maintenance requirements, making CPV systems with MJSCs less practical for widespread terrestrial use. Furthermore, outdoor environmental factors, including atmospheric variations, cloud cover, and dust accumulation, affect the spectral distribution and intensity of sunlight, leading to performance fluctuations. Another significant limitation is the thermal sensitivity of MJSCs. Unlike silicon solar cells, which demonstrate relatively better thermal stability, the efficiency of MJSCs declines at elevated temperatures, posing a challenge in regions with high ambient temperatures.

In contrast, MJSCs are the preferred choice for space applications due to their unparalleled radiation resistance, high power-to-weight ratio, and long-term stability in extreme environments. The absence of atmospheric absorption and scattering in space allows MJSCs to operate at their maximum theoretical efficiency without spectral distortions. Their ability to maintain high performance under intense solar radiation makes them ideal for satellite power systems, space probes, and extraterrestrial exploration missions [33]. Unlike terrestrial conditions, where MJSCs must contend with variable sunlight and environmental degradation, space-based MJSCs experience a relatively stable solar spectrum, optimizing their energy generation potential. Additionally, MJSCs exhibit superior radiation tolerance, allowing them to withstand prolonged exposure to cosmic radiation, which would otherwise degrade conventional photovoltaic technologies over time. Given these advantages, MJSCs continue to be the dominant technology for space power generation, where the focus is on maximizing energy output and operational longevity rather than cost reduction [148]. In space applications, independently verified record efficiencies of 46.0% under focused illumination of 508 suns for 4J and 47.1% under 143 suns under AM1.5D condition for 6J solar cell and 35.8% at AM0 (1367 $Wm^{-2}$) have previously been attained [146, 149]. For short-term space missions, power from the Si-solar cells could meet the demand, and for the long-term, III-V MJSC is the best option. The Vanguard I mission, which demonstrated the lightweight and dependability of photovoltaics in space, served as a catalyst for the adoption of space solar arrays in nearly all subsequent communication satellites, military spacecraft, and scientific space probes [150]. *Si*- and semiconductors utilized in MJSCs, viz. GaAs, InP, Ge, and related alloys (InGaP, InGaAs, InGaNAs, and AlInGaP and AlInGaAs) and InGaP/InGaAs/Ge 3JSCs and AlInGaP/AlIn-GaAs/InGaAs/Ge 4JSCs are the most commonly employed sunlight absorbers for space applications [6]. Technologies currently in development, such as Si-, thin-film-, organic-, MJ solar cells, and the Si-Quantum dot cell, can potentially achieve high *PCEs* and be applied to space applications [151]. Although MJSCs hold great promise for terrestrial applications, their high costs and technological complexities have limited their commercial deployment. However, advancements in fabrication techniques and material innovations may help bridge the cost gap, making MJSCs more accessible for large-scale energy production. The integration of MJSCs into CPV systems remains an area of active research, with ongoing efforts to enhance efficiency, improve temperature resilience, and develop cost-effective tracking solutions. While space applications will continue to dominate MJSC deployment, future breakthroughs in cost reduction and system optimization may pave the way for broader adoption in terrestrial energy markets.

## 4. Cost compatibility of MJSCs

Improving the PCE of solar cells by incorporating two or more subcells increases the overall fabrication cost. These phenomena are reflected in the techno-economic evaluation studies focused on the MJSCs [152-154] that reveal the link between the *PCE* and solar cell production cost. For instance, the fabrication cost of a 30% efficient double junction GaInP/Si and GaAs/Si solar cell account for US$ 4.85/W and US$ 7.17/W, respectively [155]. Meanwhile, the production cost of a 35% efficient 3J GaInP/GaAs/Si solar cell is US$ 8.24/W [23]. However, as mentioned earlier, incorporating additional junctions (4-, 5-, 6J) leads to increased fabrication complexity, resulting in exponential cost increases. Usually, due to the very complex fabrication procedure, MJSCs cost multiple times higher than the widely utilized *Si*-solar cells, which cost US$ 0.3 to US$ 0.35/W [155, 161] and

thin-film SC, which is US$ 1 to 1.5/W [155]. This raises relevant questions about the commercial viability of higher PCE MJSCs. Comparing power generation, a single triple-junction GaInP/GaAs/Si cell at 1000 $Wm^{-2}$ irradiation produces 350 $Wm^{-2}$ peak power. A standard *Si*-solar cell with 20% efficiency generates 200 $Wm^{-2}$. To achieve 1400 $Wm^{-2}$, four 3J GaInP/GaAs/Si cells or seven conventional *Si*-solar cells are needed. However, the considerably higher fabrication costs of 3J cells favour conventional Si solar cells [23]. Thus, MJSCs are limited to military and aerospace applications, where space and weight constraints outweigh the costs. In 2018, 3J solar cell production ranged from hundreds of kW/year to a few MW/year, comprising only 0.02% of the PV market [156]. Optimising processes and exploring alternate materials are essential to improve the performance-to-cost ratio. The manufacturing costs are US$ 100/W for 50 kW/year and US$ 70/W for 200 kW/year in production due to the underutilisation of building and fixed equipment costs [153]. High material costs, like metal-trimethyl precursors, Ge, and As, contribute to expenses. Lithography's low throughput and high costs limit large-area deposition. Transfer printing technology offers a cost-effective alternative [157]. Lower-cost techniques like electroplating are being researched to reduce contact expenses [156]. Research highlights ±20% variation effects on costs, with Ge substrate and manufacturing yield significantly impacting expenses [153]. Recycling Au reduces sensitivity to price fluctuations. The PCE of solar cells affects costs, with a 1% increase in reducing costs by US$ 2.07/W and a 1% decrease in raising costs by US$ 2.2/W [153]. Ge's high price at US$ 1,200/kg and complex extraction delay MJSC market growth. Dilute nitride antimonide sub-cells and Si substrate growth are alternatives to replace Ge [158, 159]. Si offers advantages with its larger indirect bandgap, potentially higher voltages, lighter, stronger, more abundant, and cost-effective nature [160, 162].

## 5. Discussion and outlooks

Pushing the boundaries of material design holds immense potential for enhancing the efficiency and stability of the MJSC. Exploring novel III-V alloys with wider bandgaps can enable better light harvesting across the solar spectrum, leading to more efficient photon-to-electricity conversion. Strain engineering techniques can be harnessed to meticulously manipulate bandgaps and carrier transport properties within the multijunction structure. By introducing controlled amounts of strain into specific layers, researchers can fine-tune the energy levels and improve carrier mobility, ultimately boosting overall device efficiency. Crystal defects within the material layers act as barriers for charge carriers, hindering their movement and ultimately reducing device performance. Advanced epitaxial growth techniques, such as MOVPE with precise control overgrowth parameters and post-growth treatments that can passivate defects, are essential to achieve near-defect-free material layers. This relentless pursuit of minimizing defects will pave the way for longer-lasting and more efficient solar cells. Beyond conventional architectures, innovative designs can unlock further efficiency gains and address specific application needs.

Metamaterial structures, artificially engineered materials with unconventional properties, can be incorporated into the device design to enhance light trapping and optimize light absorption within each sub-cell. These metamaterials can be meticulously designed to manipulate the path of light, ensuring that a greater portion of the incoming solar spectrum is absorbed by the active regions of the device, leading to significant efficiency improvements. Utilizing nanostructured materials like quantum wells and dots offers another exciting avenue for performance enhancement. Quantum wells, thin layers of semiconductor material with engineered bandgaps, quantum dots, and semiconductor nanoparticles with unique quantum confinement effects can improve light management and carrier confinement within the device. This can lead to higher photocurrents and, ultimately, higher efficiencies. While conventional III-V MJSCs typically utilize two or three sub-cells, stacking more than three sub-cells in a well-designed tandem configuration can enable even higher efficiencies. Each sub-cell within the tandem structure can be precisely designed to target a specific region of the solar spectrum, ensuring that a wider range of sunlight is effectively converted into electricity. This approach offers the potential to achieve record-breaking efficiencies, pushing the boundaries of solar energy conversion.

Achieving cost-effective and scalable fabrication processes is paramount for transforming III-V MJSCs from research sights to commercially viable products. Critical research efforts are needed to bridge the gap between laboratory demonstrations and large-scale production. III-V MJSCs are typically grown on expensive III-V substrates like gallium arsenide. Utilizing alternative substrates like germanium or silicon carbide (SiC) offers a promising route to reduce production costs. However, significant challenges arise due to lattice mismatch between the epitaxial layer and the substrate itself. This mismatch can lead to defects and reduced device performance. Further research on advanced epitaxial growth techniques and innovative buffer layer designs is crucial to overcome these challenges and unlock the cost-reduction potential of alternative substrates. Developing high-throughput deposition techniques like roll-to-roll processing for III-V materials is crucial for mass production and cost reduction. Currently, epitaxial growth techniques like MOVPE are often batch processes, meaning they can only grow a limited number of layers at a time. Roll-to-roll processing, on the other hand, allows for continuous deposition of thin films on a moving substrate, enabling large-scale production and significantly reducing costs. However, adapting these techniques for III-V materials presents challenges due to their higher growth temperatures and sensitivity to process variations. Further research in this area is essential for making III-V MJSCs a commercially viable option.

Extensive research on long-term stability and degradation mechanisms under real-world operating conditions is vital for ensuring reliable and durable solar cells. III-V MJSCs, despite their impressive efficiencies, can be susceptible to degradation over time due to factors like exposure to sunlight, heat, and humidity. Understanding these degradation mechanisms is crucial for developing strategies to improve device lifetime. This may involve exploring new material combinations that are more resistant to degradation, developing encapsulation techniques that protect the device from harsh environmental factors, and optimizing device design to minimize internal stresses. By addressing these issues, researchers can build III-V MJSCs that can maintain their high performance over extended periods, ensuring a valuable return on investment for users.

Developing sophisticated in-situ and innovative specific characterization tools is essential for gaining deeper insights into material and device behaviour. These tools can help optimize growth parameters, identify performance limitations, and guide the development of new materials and device architectures. In-situ characterization techniques allow researchers to monitor the growth process of III-V materials in real time. This provides valuable information about how growth parameters like temperature, pressure, and precursor flow rate influence the properties of the deposited layers. By understanding these relationships, researchers can fine-tune the growth process to achieve high-quality materials with optimal properties for efficient solar cell operation. Operando characterization techniques allow researchers to probe the behaviour of III-V MJSCs under actual operating conditions, such as under illumination and at elevated temperatures. This provides insights into carrier transport mechanisms, recombination processes, and other factors influencing device performance. By analyzing the data obtained from operando characterization, researchers can identify performance limitations and develop strategies to address them, ultimately leading to higher efficiency and improved device performance.

The development of accurate computational models in MJSC technology is essential, acting as a bridge between theoretical understanding and experimental validation. These models must incorporate realistic representations of material imperfections, such as point defects and dislocations, which significantly impact carrier generation and recombination dynamics. Additionally, the models should account for non-ideal carrier transport mechanisms, including trap-assisted tunnelling and interfacial recombination across different junctions. Effective modelling of optical phenomena—such as wavelength-dependent absorption and parasitic absorption—is crucial for realistic efficiency predictions, especially when implementing light management strategies like anti-reflection coatings and textured surfaces. To improve predictive accuracy, it is important to integrate experimental calibration data, including temperature-dependent material parameters and interface recombination velocities. Well-calibrated models can guide material selection, optimize layer thicknesses, and identify performance-limiting factors in existing devices. Continued refinement of these models through experimental validation is vital for advancing theoretical knowledge and practical applications in MJSCs. Incorporating emerging phenomena—such as quantum confinement effects and advanced carrier multiplication—will enhance the ability to design and optimize next-generation MJSCs, thereby reducing experimental development time and resources.

Exploring synergies between III-V MJSCs and other technologies can create innovative solutions that push the boundaries of solar energy applications. Integrating III-V MJSCs with CPV systems offers a promising approach for achieving ultra-high efficiencies. CPV systems utilize lenses or mirrors to concentrate sunlight onto a smaller area, allowing for the use of smaller and potentially more expensive solar cells like III-V MJSCs. This approach can be particularly beneficial for applications where space is limited, such as rooftops or utility-scale solar farms. Further research on optimizing the integration between III-V MJSCs and CPV systems can unlock new possibilities for high-efficiency solar power generation. Tailoring III-V MJSCs for space applications requires addressing specific challenges like radiation resistance, extreme temperature variations, and high vacuum environments. These harsh environments can significantly impact the performance and lifespan of solar cells. Research efforts are underway to develop radiation-resistant materials, robust device architectures, and advanced encapsulation techniques to ensure the reliable operation of III-V MJSCs in space. With further advancements, III-V MJSCs have the potential to become the dominant technology for powering satellites, spacecraft, and future lunar or Martian vehicles.

The journey of III-V MJSCs towards commercial viability and widespread market adoption is fraught with challenges yet sustained by immense potential. Addressing the key commercial viability issues is a matter of technological innovation, strategic market positioning, and regulatory navigation. Future advancements depend significantly on interdisciplinary collaboration, where insights from material science, manufacturing technology, market analysis, and regulatory expertise converge. Efforts to enhance these solar cells' economic and operational feasibility will pave the way for their successful integration into the global energy market. Such integration promises to fulfil niche market needs and propel these technologies into mainstream applications, contributing significantly to global renewable energy targets. Continued investment in research and development, coupled with a proactive approach to addressing market and regulatory challenges, will be crucial for unlocking the full potential of III-V MJSCs. This concerted effort will ensure that these advanced photovoltaic systems move from laboratory settings to commercial rooftops and beyond, helping to meet the urgent demand for sustainable and efficient energy solutions worldwide.

## 6. Conclusion

III-V compound semiconductor multijunction solar cells are champions in achieving ultra-high conversion efficiencies, surpassing the Shockley-Queisser limit. While they offer about double the efficiency of first-generation Si solar cells and can be customized for various applications, commercial adoption still faces challenges. Therefore, people are trying to further enhance PCE efficiency by incorporating new materials, models, and concepts, minimizing thermalization and transmission losses. In this work, we critically reviewed several essential aspects that were not previously addressed, including:

- The advancement of subcell combinations from two to six junctions, considering the progress of incorporating semiconductor materials, ultimately enhanced the PCE up to 47.1%.
- Several computational and experimental approaches have been addressed to advance these solar cell developments. The in-house code was frequently utilized for most of the numerical simulations, whereas MBE is found to be an effective technique for III-V MJSC fabrications.
- Different semiconductor materials employed as the subcells of MJSCs, such as III-V, hybrid tandem, organic-inorganic PSCs, and a few emerging materials, have been investigated.

The fabrication process of MJSCs is complex, and the scarcity of critical materials combined with slow growth rates result in low throughput, making it difficult for large-scale commercialization. The integration of CPV technology could enhance photovoltaic performance and reduce manufacturing costs. To surpass other photovoltaic technologies, the manufacturing costs of the MJSCs must decrease significantly. However, advancing the MJSC concept by integrating PSCs with Si-solar cells could lead to the commercialization of this technology. This can be achieved by utilizing full perovskite, III-V/perovskite, or perovskite/Si tandem MJSCs in the future.

## Acknowledgments

This work was funded by the Bangladesh Council of Scientific and Industrial Research (BCSIR) under the R&D grant Ref. 39.02.0000.011.14.169.2023/877/IERD-9.

## Data Availability

Associated data are available on reasonable request.

## Conflict of Interest

Authors declare there is no conflict of interest.